\documentclass{amsproc}

\usepackage{amssymb}

\usepackage{simplewick,hyperref}

\newtheorem{theorem}{Theorem}[section]

\theoremstyle{definition}

\def\ed{\, \textrm{d}}
\def\FF{\mathcal F}
\def\bx{\mathbf{x}}
\def\bX{\mathbf{X}}

\def\bP{\mathbf{P}}

\def\half{\frac{1}{2}}
\newcommand{\OO}{\mathcal{O}}

\newcommand{\SU}{\mathrm{SU}}

\newcommand{\cA}{\mathcal A}

\newcommand{\cM}{\mathcal M}

\newcommand{\pd}{\partial}

\theoremstyle{remark}

\numberwithin{equation}{section}

\begin{document}

  \rightline{QMUL-PH-14-07}
  \rightline{MIFPA-14-25}

   \vspace{1 truecm}

\title{Instanton-soliton loops in 5D super-Yang-Mills}


\author[C.~Papageorgakis]{Constantinos Papageorgakis}
\address{NHETC and Department of Physics and Astronomy, Rutgers, The State
  University of New Jersey, Piscataway, NJ 08854-8019, USA}
\address{CRST and School of Physics and Astronomy, Queen Mary
  University of London, E1 4NS, UK}
\curraddr{}
\email{c.papageorgakis@qmul.ac.uk}
\thanks{C.P. is a Royal Society Research Fellow and is partly supported
  by U.S. DOE Grants DOE-SC0010008, DOE-ARRA-SC0003883
  and DOE-DE-SC000789.}

\author[A.~B.~Royston]{Andrew B.~Royston}
\address{George P.~\& Cynthia Woods Mitchell Institute for Fundamental
Physics and Astronomy, Texas A\& M University, College Station, TX
77843, USA}
\curraddr{}
\email{aroyston@physics.tamu.edu}
\thanks{}


\date{}

\begin{abstract}
  Soliton contributions to perturbative processes in QFT are
  controlled by a form factor, which depends on the soliton size.  We
  provide a demonstration of this fact in a class of scalar theories
  with generic moduli spaces. We then argue that for
  instanton-solitons in 5D super-Yang-Mills theory the analogous form
  factor does not lead to faster-than-any-power suppression in the
  perturbative coupling.  We also discuss the implications of such
  contributions for the UV behavior of maximally supersymmetric
  Yang-Mills in 5D and its relation to the (2,0) CFT in 6D. This is a
  contribution to the proceedings of the ``String Math 2013''
  conference and is a condensed version of results appearing in
  \cite{Papageorgakis:2014dma,Papageorgakis:2014jia}.
\end{abstract}

\maketitle


\section{5D MSYM and the (2,0) SCFT in 6D}

Over the last few years, there has been renewed interest in the
relation between the maximally supersymmetric Yang-Mills theory (MSYM)
in 5D and the nonabelian (2,0) tensor CFT in 6D. In
\cite{Douglas:2010iu,Lambert:2010iw} it was conjectured that 5D MSYM
with coupling $g_{\rm YM}$ is exactly the (2,0) theory on a circle of
radius
\begin{equation}
R= \frac{g_{\rm YM}^2}{4\pi}\;.
\end{equation}
This relies on the observation that Kaluza-Klein momentum along the
circle can be identified with instanton charge in the 5D theory; the
latter is a topological charge carried by soliton configurations
\cite{Rozali:1997cb, Berkooz:1997cq}. Equivalently, in the strong
coupling limit 5D MSYM should define the fully decompactified (2,0)
theory, which in turn is expected to describe the low-energy dynamics
of multiple M5-branes \cite{Witten:1995zh, Strominger:1995ac}.

At the same time, since gauge theories above four dimensions are
power-counting nonrenormalizable, one would expect that 5D MSYM should
be treated as an effective theory in the Wilsonian sense, that is only
defined up to some cutoff scale. It is then reasonable to wonder what
it means to consider such a theory at strong coupling (and hence high
energies) and how that would give rise to the well-defined (2,0) CFT.

This tension naturally leads to revisiting the UV behavior of 5D MSYM
in the context of perturbative renormalization. Despite $\mathcal N
=2$ supersymmetry being responsible for the absence of UV divergences
at low orders \cite{Howe:2002ui}, the first logarithmic divergence was
explicitly seen at 6 loops in \cite{Bern:2012di}. However, instead of
immediately taking the cutoff to infinity and declaring the theory
UV-divergent, one has to also investigate possible contributions
associated with virtual soliton states. The simplest such contribution
can be related, via the optical theorem, to the soliton-antisoliton
pair production amplitude. The latter is conventionally believed to be
``exponentially suppressed'' and as a result soliton loops are usually
ignored.

Here we will argue that instanton-soliton pair production in 5D MSYM
does not fall faster than any power in the effective dimensionless
coupling controlling a given perturbative process
\cite{Papageorgakis:2014dma}. Motivated by \cite{Douglas:2010iu,
  Lambert:2010iw}, one could then envision a mechanism through which
soliton loops would lead to exact cancelations against the
perturbative UV divergences and render the theory well defined. We
stress that we are not treating 5D MSYM as an effective theory in the
Wilsonian sense, and instead we are supposing that the Lagrangian
provides a microscopic definition of the theory. This point of view
can only make sense if the theory is finite.

\section{Soliton pair production and form factors}

According to standard QFT lore, soliton production is ``exponentially
suppressed'' at small coupling and hence unimportant for perturbative
physics. However, upon careful consideration one can formulate a more
refined version of that statement: the exponential dependence should
appear in the dimensionless ratio of the soliton size ($R_S$) over its
Compton wavelength ($R_C$), $e^{-R_S/R_C}$
\cite{Banks:2012dp,Papageorgakis:2014dma}. On the one hand, when the
size is fixed to a value much larger than the Compton wavelength one
recovers the expected suppression, as e.g.~for `t Hooft--Polyakov
monopoles in Yang--Mills--Higgs theory. On the other, for situations
where the size is a modulus that ranges over values on the order of
the Compton wavelength, one might expect that small solitons would not
be suppressed compared to perturbative processes. As we will see, this
is precisely the setup for instanton-solitons in five dimensions.

To that end, it is instructive to revisit the derivation of the
soliton form factor for the simple case of scalar theories
\cite{Papageorgakis:2014dma}. Consider the following class of
Lagrangians
\begin{equation}\label{lagrangian}
L = \frac{1}{g^2}\int \ed\bx \left\{ \half \dot{\Phi} \cdot \dot{\Phi}
  -\half\pd_\bx \Phi \cdot \pd_\bx \Phi - V(\Phi) \right\}~.
\end{equation}
We denote by $\bx$ a $(D-1)$-dimensional position vector, while
$\ed\bx$ is shorthand for $\ed^{D-1} x$. The field $\Phi$ is $\mathbb{R}^n$-valued
and $\cdot$ denotes the Euclidean dot product.  Here we assume that
the potential has a dimensionless parameter $g$ controlling the
perturbative expansion. Then, in terms of canonically normalized
fields $\tilde{\Phi} = g^{-1} \Phi$, we have
$\tilde{V}(\tilde{\Phi};g) = g^{-2} \tilde{V}(g\tilde{\Phi};1)$, while
we have also set $V(\Phi) = \tilde{V}(g \tilde{\Phi};1)$
\cite{Goldstone:1974gf}.

We are interested in soliton solutions, classically described by
localized, finite-energy field configurations and denoted by
$\phi$. Such classical solutions for a fixed topological
sector\footnote{The sectors are labeled by homotopy equivalence
  classes $\pi_{D-2}(M_{\rm vac})$, where $M_{\rm vac} := \{ \Phi ~|~
  V(\Phi) = 0 \}$.  See e.g.~\cite{MR2068924}. Note that
  although Derrick's theorem \cite{Derrick:1964ww,MR2068924} precludes
  the existence of soliton solutions for $D > 2$ in the class of
  linear sigma models considered here, it is no more difficult to
  leave $D$ arbitrary.}  usually come in a smooth family parameterized
by a collection of moduli, $U^M$, where $M = i,m$. A subset of these
moduli always consist of the center-of-mass positions, $ U^i = \bX$;
we will call their conjugate momenta $\bP$. $U^m$ then parameterize
all remaining ``centered'' moduli. We denote the moduli space of
soliton solutions for a given fixed topological charge as $\mathcal
M$; it represents a local minimum of the energy functional.

In the presence of a soliton a new sector of the quantum theory opens
up, which is orthogonal to the vacuum sector since solitons carry a
conserved topological charge \cite{Goldstone:1974gf}. Single particle
states in the soliton sector form a subspace of the total
single-particle Hilbert space and one can study processes involving
both perturbative particles and solitons as asymptotic states. Such
soliton states can be chosen to be momentum eigenstates,
$|\bf{P}\rangle$. Note that, apart from the soliton's actual energy
and momenta, these states can carry extra labels corresponding to
eigenvalues of additional operators that commute with the
Hamiltonian. These depend on the particulars of the theory and will be
left implicit for the rest of our discussion.

Let us now study the soliton pair-production amplitude, involving a
perturbative incoming excitation of (off-shell) momentum $k$ and a
soliton-antisoliton outgoing pair of (on-shell) momenta $P_f$ and
$-P_i$ respectively.  Although it is unclear how one should proceed
directly---since there exists no known associated analytic classical
solution and hence no semiclassical expansion scheme---crossing
symmetry relates soliton pair-production to a process where the
soliton (baryon) absorbs the perturbative excitation (meson)
\begin{equation}\label{crossing} 
  \cA(k\to P_f,-\bar P_i) = \cA (P_i, k\to P_f)\;.
\end{equation} 
Note that this is an equality between amplitudes in distinct
topological sectors. The RHS is related to the form factor
\begin{align}\label{amplitude}
   i (2\pi)^D \delta^{(D)}(k+P_i-P_f)\cA &(P_i,k\to P_f)
  =\cr =& \int \ed^Dx \;e^{-i k\cdot x}\langle \bP_f|{\rm T}\big\{  \Phi(x)  \;e^{-i \int \ed t'
    H_I(t')}\big\} |\bP_i \rangle\;, \quad
\end{align}
where $H_I$ denotes the interaction Hamiltonian.

The Hamiltonian obtained from  \eqref{lagrangian} is simply
\begin{equation}
  H = \int \ed\bx \left\{ \frac{g^2}{2} \Pi \cdot \Pi
    +\frac{1}{g^2}\left(\half\pd_\bx \Phi \cdot \pd_\bx \Phi +
      V(\Phi)\right) \right\}\;. 
\end{equation}
Through a canonical transformation \cite{Papageorgakis:2014jia} the
original conjugate pair of variables $(\Phi,\Pi)$ can be related to
the new pairs $(U^M, p_N)$, $(\chi,\pi)$ capturing the collective
coordinate and massive oscillator dynamics respectively. We have
\begin{align}
  \Phi(x) &= \phi(\bx;U) + g \;\chi(x;U) \cr
  \Pi(x) &= \half \left( a^M \pd_M \phi(\bx;U) + \pd_M \phi(\bx;U)
    \bar{a}^M \right) + \frac{1}{g} \;\pi(x;U) \;, 
\end{align}
subject to the constraints
\begin{equation}\label{constraints}
 F_{1,M}:=\int \ed\bx\; \chi\cdot \pd_M \phi =0\;,\qquad
   F_{2,M}:=\int \ed\bx \;\pi\cdot \pd_M \phi = 0\;,
\end{equation}
which ensure that the fluctuations $\chi,\pi$ are orthogonal to the
zero-modes $\pd_M \phi$.  Here we have inserted factors of $g$ so that
the fluctuation fields are canonically normalized.  The functionals
$a^M, \bar{a}^M$ are given by
\begin{equation}
  a^N = \frac{1}{g^2}\left (p_M -\int \pi \cdot \pd_M
    \chi \right) C^{MN}~, \quad \bar{a}^M = \frac{1}{g^2} C^{MN} \left (p_M -
    \int \pd_M \chi  \cdot \pi \right)~, 
\end{equation}
where $C = (G - g \Xi)^{-1}$ with
\begin{equation}\label{xig}
  \Xi_{MN} = \frac{1}{g^2}  \int\chi \cdot\pd_M \pd_N \phi\;,\qquad
  G_{MN} = \frac{1}{g^2}  \int
  \pd_M\phi\cdot \pd_N \phi ~.
\end{equation}
Here $G_{MN}$ is the metric on moduli space, induced from the flat metric
on field configuration space.

In terms of these new variables the Hamiltonian can be written as
\begin{align}
  \label{eq:3}
H & =   \frac{g^4}{2}  a^M
 G_{MN}  a^N  + {\rm v}(U^m)  +\int \Big[\half
   \pi \cdot \pi + g\; s\cdot  \chi + 
\frac{1}{2}  \chi \cdot \Delta  \chi + V_I(
\chi)  \Big]+ \cr
& \qquad \qquad \qquad \qquad \qquad \qquad \qquad \qquad \qquad \qquad \qquad  + \OO(g^2) ~, 
\end{align}
with $V_I(\chi)$ denoting cubic and higher-order interaction terms in
the fluctuations $\chi$ coming from the original potential. In writing
the above, we have ignored operator-ordering ambiguities, such that
$a^M = \bar a^M + \OO(g^2)$. These corrections correspond to two-loop
effects that will not be important for the rest of our calculation.

We have also defined
\begin{align}\label{sourceOp}
  & s(\bx;U^m) :=\frac{1}{g^2} \Big( - \pd_{\bx}^2 \phi + \frac{\pd V}{\pd \Phi}
  \bigg|_{\Phi = \phi} \Big)~, \qquad
  \Delta  := -\delta_{ab} \pd_{\bx}^2 + \frac{
  \delta^2 V}{\delta \Phi \delta\Phi}\bigg|_{\Phi = \phi}~, \cr 
 & {\rm v}(U^m)  := \frac{1}{g^2} \int \ed\bx \left(
    \half \pd_{\bx} \phi 
    \cdot \pd_{\bx} \phi + V(\phi) \right) = M_{\rm cl} + \delta
  {\rm v}(U^m)\;.
\end{align}
If $\phi$ is an exact solution to the time-independent equations of motion then $s(\bx;U^m) = 0$ and $\delta {\rm v}(U^m) = 0$.  However, in theories with centered moduli it is sometimes convenient to expand around a configuration that is only an approximate solution.

The regime needed to extract information about the pair-creation
process through crossing symmetry requires large velocity exchange and
hence momentum transfer of the order of the soliton mass, $\bP\sim
\frac{m}{g^2}$ with $m$ the meson mass. Therefore, the conventional
small-velocity (Manton) approximation usually implemented in the literature is
not sufficient for our purposes.

\section{Relativistic scalar form factor}\label{scalar}

In the case of the two-dimensional kink in $\Phi^4$ theory, seminal
work by Gervais, Jevicki and Sakita \cite{Gervais:1975pa} showed how
velocity corrections can be systematically accounted for, to recover
the covariant expression for the soliton energy, $M_{\rm cl}\to \sqrt{
  {\bf P}^2 + M_{\rm cl}^2}$. This answer is to be expected, since the
starting point is a Lorentz-invariant theory. We will now show how the
same techniques can be applied in the more general class of
Lorentz-invariant theories considered here. We will be interested in
evaluating the form factor \eqref{amplitude} rather than the soliton
energy. Fortunately, the techniques of \cite{Gervais:1975pa} have been adapted to this
context by \cite{Dorey:1993by}, the methodology of which we will be
following closely.

The two qualitative differences between the general case and the kink
in $\Phi^4$ theory are: a) lack of an explicit classical soliton
solution to work with and b) the possible presence of centered
moduli. Both can be taken into account and their discussion can be
appropriately modified, provided we continue to make the simplifying
assumptions of the Manton (small-velocity and small
moduli-space-potential) approximation for the dynamics of the centered
moduli.  Specifically, we will impose $p_m/m \sim \OO(1/g)$ and
$s(\bx;U^m) \sim \OO(1)$, but we will assume $p_i/m = \bP/m \sim
\OO(1/g^2)$.

The transition amplitude from an initial state $i$ described by the functional $\Psi_i(U^M(-T);\chi)$ to a final state $f$ described by the functional $\Psi_f(U^M(T);\chi)$ is
\begin{align}\label{transamp}
S_{fi} & =~ \int [DU Dp D\chi D\pi] \delta(F_1) \delta(F_2) e^{i \int_{-T}^T \ed t L} \Psi_{f}^\ast \Psi_i~, \quad \textrm{with} \cr
L & =~p_M \dot{U}^M + \int \ed \bx \;\pi \dot{\chi} - H~.
\end{align}
An incoming soliton state of momentum $\bP_i$ is defined by taking
$\Psi_i = e^{i \bP_i \cdot \bX_i}\tilde \Psi_{i}(U^m)$, where $\bX_i =
\bX(-T)$, and similarly for outgoing soliton states. The $\tilde
\Psi_{i,f}$ are wavefunctions on the centered moduli space. In general
we will denote quantities associated with the centered part of the
moduli space with a tilde. We can consider time-ordered correlators of
the meson field between soliton states by inserting appropriate
factors of $\Phi(x_1)\cdots \Phi(x_n)$ under the path integral, and
using the relation $\Phi(x) = \phi(\bx - \bX(t);U^m) + g \chi(t,\bx -
\bX(t);U^m)$.

We are interested in the particular case of the 1-point function and
hence in 
\begin{align}\label{full1pt}
\langle \bP_f,T | \Phi(x) | \bP_i,-T \rangle =&~ \int [D\bX D \bP] e^{i
  (\bX_i \cdot \bP_i - \bX_f \cdot\bP_f)} \int[DU^m Dp_n] \tilde \Psi^*_f\tilde \Psi_i \times \cr
&~ \times \int [D \chi D \pi] \delta(F_1) \delta(F_2) e^{i \int_{-T}^T \ed t L} \Phi[U,p;\chi](x)~.
\end{align}

Let us focus first on the internal path integral over $\chi$ and $\pi$
for which we will proceed to compute the leading contribution at small
$g$. This was done in \cite{Gervais:1975pa} for the case of the
0-point function by evaluating the action on the saddle-point solution
for $\chi,\pi$ corresponding to the moving soliton. One can argue
\cite{Dorey:1993by,Papageorgakis:2014dma} that the same saddle point
solution gives the leading contribution to the 1-point function, even
though one should now be solving the equations of motion with source.
This is a special feature of working with the 1-point function and
would not be true for higher-point functions. We denote this saddle
point $(\chi_{\rm cl}, \pi_{\rm cl})$ and expand the fields as $\chi =
\chi_{\rm cl} + \delta\chi$, $\pi = \pi_{\rm cl} + \delta \pi$.

Starting with the Hamiltonian \eqref{eq:3} one can find a saddle-point
solution to the $\chi,\pi$ equations of motion perturbatively in $g$
by making use of the above-mentioned scaling assumptions for the
coordinate momenta \cite{Papageorgakis:2014dma}. One finds
\begin{equation}\label{chicl}
\chi_{\rm cl} = g^{-1} \phi\left( \Lambda (\bx -
  \bX); U^m \right) - g^{-1} \phi\left((\bx -
  \bX); U^m \right) + \OO(1)~,
\end{equation}
where 
\begin{equation}\label{chisol2}
{\Lambda^i}_j = {\delta^i}_j + \left( \sqrt{1 + \frac{\bP^2}{M_{\rm cl}^2}} - 1 \right) \frac{p^i p_j}{\bP^2} ~
\end{equation}
is a Lorentz contraction factor. The insertion can then be expressed
as
\begin{align}
  \label{eq:9}
  \Phi  =   \phi\left( \Lambda(\bx - \bX); U^m \right)+ \OO(g)
  \equiv \Phi_{\rm cl} + \OO(g)\;.
\end{align}

With this solution in hand, we want to evaluate \eqref{full1pt} in the
presence of centered moduli. For this, we also need the Lagrangian
evaluated on the solution:
\begin{equation}\label{fullL}
L = \bP \cdot \dot{\bX} - \sqrt{\bP^2 + M_{\rm cl}^2} + L^{(0)}[U^m, p_m; \delta\chi,\delta\pi; \bP] + L_{\rm int}~,
\end{equation}
where $L_{\rm int}$ starts at $\OO(g)$ and
\begin{equation}\label{laglag}
L^{(0)} = p_m \dot{U}^m - \tilde H_{\rm eff}[U^m,p_m;\bP]
\end{equation}
is an $\OO(1)$ contribution describing the dynamics of the centered
moduli, whose precise form we will not require. $\tilde H_{\rm eff}$
includes the 1-loop potential from integrating out the fluctuation
fields $(\delta\chi,\delta\pi)$. The leading contribution to
\eqref{full1pt} then takes the form
\begin{align}
  & \langle \bP_f | \Phi(x) | \bP_i \rangle = \int [D \bX D\bP] e^{i
    (\bX_i \cdot \bP_i - \bX_f\cdot \bP_f)} e^{i \int \ed t (\bP\cdot  \dot{\bX} -
    \sqrt{ \bP^2 + M_{\rm cl}^2} ) } \times \cr &\times \int_{\tilde
    \cM} \ed U\sqrt{\tilde G}\; \tilde\Psi^*_f(U^m;\bP) \Phi_{\rm
    cl}[\bX,\bP;U^m]\tilde\Psi_i(U^m;\bP) \left(1 + \OO(g)\right) ~.
\end{align}
In the above we have expressed the centered moduli space path integral
as a position-basis matrix element in the quantum mechanics on the
centered moduli space with Hamiltonian $\tilde H_{\rm eff}$. Note that
the $(\bX,\bP)$ path integral is a functional integral representation
of the quantum mechanics for a relativistic particle. From the point
of view of the translational moduli space dynamics, $U^m$ are merely
parameters, so we can carry out the functional integration over $\bX$
and $\bP$ first and then integrate over the centered moduli space.

This was carried out in \cite{Papageorgakis:2014dma} employing the
techniques of \cite{Dorey:1993by}. Using that result, which is
specific to two dimensions, we find that the amplitude
\eqref{amplitude} takes the following form to leading order:
\begin{align}
  \label{eq:10}
  \cA(P_i, k \to P_f) \sim&~ \int_{\tilde \cM} \ed U\sqrt{\tilde
    G}\;\tilde\Psi_f^* \FF[ \phi] \left(
    \frac{2R_S(U^m)}{R_C}\zeta(P_f, P_i) \right) \tilde\Psi_i\;,
\end{align}
where $\FF[\phi](u) =\int \ed v\;  e^{- i u v} \phi(v)$ is the Fourier transform of the classical
soliton profile, $  \tilde\Psi_{i,f} = \tilde\Psi_{i,f}
(U^m;P_{i,f})$ and
\begin{align}
  \zeta(P_f, P_i) := \frac{2 \epsilon_{\mu\nu}P_f^\mu P^\nu_i}{(P_f + P_i)^2}\;.
\end{align}
The quantity $R_S(U^m)$, inserted on dimensional grounds,
characterizes the size of the soliton. For example, in $\Phi^4$ theory
$R_S = 1/m$, with $m$ the meson mass.  As we previously indicated, in
the general class of theories considered here it can in principle be a
function of the centered moduli. $R_C = 1/M_{\rm cl}$ is the soliton
Compton wavelength.

Now, given that the classical soliton profile $\phi$ is a smooth
$(C^\infty)$ function of $\bx - \bX$, we can draw a rather strong
conclusion about the asymptotic behavior of the Fourier transform in
\eqref{eq:10}. For any values of momenta such that $\zeta$ is not 
$O(g^2)$
or smaller, it is the $2R_S/R_C$ factor that controls the parametric
size of the argument of the Fourier transform. Given this, and as long
as the soliton size is bounded away from zero, $R_S^{\rm min}>0$, we
will have that $(2R_S/R_C)|\zeta|\to\infty$ in the semiclassical
limit. The Riemann--Lebesgue lemma then implies that\footnote{As
  stated by the Riemann-Lebesgue lemma, the Fourier transform
  $\mathcal F[f](p)$ of an $L^1$-function $f(x)$ goes to zero as $|p|
  \to \infty$. Accordingly, if $f(x)$ is $C^\infty$, $\mathcal
  F[f^{(n)}](p) = (ip)^n \mathcal F[f](p)$ should also go to zero as
  $p\to \infty$; i.e. $\mathcal F[f](p)$ goes to zero faster than any
  power.}
\begin{equation}\label{Riemann}
\lim_{g\to 0} \mathcal F[\phi ]\left(\frac{2R_S(U^m)}{R_C} \zeta\right)\sim  \;e^{- \frac{2R_S(U^m)}{R_C} |\zeta|} \;.
\end{equation}
Let us emphasize that the exponential on the RHS is a typical function
exhibiting a faster-than-any-power falloff that we use for
concreteness, but the exact expression will depend on the details of
the theory under consideration. In any case, the important property
for our purposes is the faster-than-any-power falloff.

This leads to the expression
\begin{align}
  \label{eq:14}
  \cA(P_i,k \to P_f) \sim& \int_{\tilde \cM} \ed U\sqrt{\tilde
    G}\;\tilde\Psi_f^* e^{- \frac{2R_S(U^m)}{R_C} |\zeta|} \tilde\Psi_i
\end{align}
for the leading contribution to the form factor as $g\to 0$. Note that the centered moduli space represents the internal degrees of
freedom of the single-particle state. A field theory interpretation
requires a single-particle state to have a finite number of internal
degrees of freedom. The eigenvalues labeling these internal degrees of
freedom should be discrete eigenvalues of the centered-moduli-space
Hamiltonian $\tilde H_{\rm eff}$.  Hence the wavefunctions on the
centered moduli space $\tilde{\Psi}$ should be $L^2$; this is
  automatically the case if $\tilde{\mathcal M}$ is compact. Then we
  have the inequalities
  \begin{align}
    \int_{\tilde \cM} \ed U\sqrt{\tilde G}\;\tilde\Psi_f^* e^{-
      \frac{2R_S(U^m)}{R_C} |\zeta|} \tilde\Psi_i & \le \int_{\tilde
      \cM} \ed U\sqrt{\tilde G}\;|\tilde\Psi_f^* \tilde\Psi_i | e^{-
      \frac{2R_S(U^m)}{R_C} |\zeta|} \cr & \le e^{- \frac{2R_S^{\rm
          min}}{R_C} |\zeta|} ||\tilde\Psi_f^* \tilde\Psi_i
    ||_{L^1}\cr & \le e^{- \frac{2R_S^{\rm min}}{R_C} |\zeta|}
    ||\tilde\Psi_f||_{L^2} ||\tilde\Psi_i ||_{L^2}\cr & = e^{-
      \frac{2R_S^{\rm min}}{R_C} |\zeta|} \;,
  \end{align}
where in the second-last step we used H\"older's inequality. Hence we have reached the result
\begin{align}
  \cA(P_i,k \to P_f) \lesssim e^{- \frac{2R_S^{\rm min}}{R_C} |\zeta|}\;. 
\end{align}

In order to use this result to obtain the pair-production amplitude
using crossing symmetry, we first re-write $\zeta$ in terms of $k^2 =
(P_f - P_i)^2$. Making use of the fact that $P_{i,f}$ are on shell,
one can show
\begin{align}
  \label{eq:20}
  \zeta = \sqrt{\frac{k^2}{k^2 - 4 M_{\rm cl}^2}}\;.
\end{align}
The above result is consistent with expectations. First, on physical
grounds the form factor should be a function of the momentum transfer
only; $\zeta = \zeta(k^2)$. Second, as $k^2\to \infty$ we expect
$\zeta(k^2)\to O(1)$; otherwise, one would obtain an amplitude with
exponential behavior for large $k^2$, in contradiction with the
large-momentum behavior of asymptotically free theories. Finally, it
agrees exactly with the prescription proposed in \cite{Ji:1991ff} in
the context of the Skyrme model, where it was also observed that
exponential falloff is inconsistent with asymptotic freedom. 

This quantity can be analytically continued from spacelike to timelike
$k^2$ and goes to the same value as $k^2\to \infty$ in any direction
on the complex plane. Thus, via crossing symmetry we arrive at
\begin{align}\label{pairprod}
    \cA(k \to -P_i, P_f) \lesssim e^{- \frac{2R_S^{\rm min}}{R_C} \zeta}\;.
\end{align}
This is in agreement with the original expectation from dimensional
analysis. Note that if $R_S^{\rm min}$ is of order $R_C$ this does not lead to
suppression.

\section{Instanton-soliton loops in 5D MSYM}\label{instantons}

Let us now apply the above result to the case of interest,
i.e. instanton-solitons in 5D MSYM. Instanton-solitons are
finite-energy $\half$-BPS configurations, obtained by solving the
selfduality equation for the gauge field strength in the four spatial
directions, $F = \star_4 F$, and have mass $M_{\rm cl} \propto
1/g_{\rm YM}^2$. As such, they are described by standard 4D instanton
solutions, which for topological charge $c_2(F) = 1$ and $\SU(2)$
gauge group, correspond to classical gauge fields given by
\begin{equation}
  A_i = U(\vec
  \theta)^{-1}\Big(\frac{\eta_{ij}^a(\bx-\bX)^j}{((\bx-\bX)^2 +
    \rho^2)}T^a\Big)U(\vec\theta)\;,\qquad A_0 = 0\;,
\end{equation}
with $a = 1,2,3$, $i = 1,...,4$ and $\eta_{ij}^a$ the `t Hooft
symbols. This solution has eight moduli: four center-of-mass
collective coordinates $\bX$, a size modulus $\rho$ and three Euler
angles $\vec \theta$ parameterizing global gauge transformations. The
associated moduli space is a hyperk\"ahler manifold
\begin{equation}
 \mathcal M = \mathbb R^4 \times \mathbb R_+ \times S^3/\mathbb Z_2\;,
\end{equation}
with metric
\begin{equation}
 \ed s^2 = \frac{4\pi^2}{g_{\rm YM}^2}\Big[\delta_{ij}\ed \bX^i \ed\bX^j +
 2(\ed\rho^2+ \rho^2 \tilde G_{\alpha\beta}\ed\theta^\alpha \ed\theta^\beta)\Big] \;,
\end{equation}
where $\tilde G_{\alpha\beta}$ is the metric on ${\rm SO}(3) \cong
S^3/\mathbb Z_2$, the group of effective global gauge transformations.

The existence of the noncompact size modulus $\rho$ means that we can
have arbitrarily small or large soliton sizes.  However, it is also
responsible for the absence of $L^2$-normalizable wavefunctions on the
centered moduli space $\tilde\cM $. This renders the interpretation of
instanton-solitons as asymptotic states confusing, since they would
correspond to particles with a continuous infinity of internal degrees
of freedom.

Moreover, the semiclassical expansion parameter in this theory is in
fact $g^2 = g_{\rm YM}^2/\rho$, which coincides with $R_C/R_S$. In
particular, note that $g(\rho)$ is moduli dependent. In the context of
finding the saddle-point solution \eqref{chicl} we can imagine a fixed
$\rho$, such that $g(\rho)$ is small. However, when evaluating
amplitudes, where one must integrate over all sizes, the semiclassical
approximation breaks down. Consequently, the small-sized
instanton-solitons invalidate our argument for exponential
suppression.

One can attempt to circumvent this conclusion by turning on a scalar
VEV, $\langle \Phi \rangle \neq 0$, and going out onto the Coulomb
branch.\footnote{Here $\Phi$ is one of the five adjoint scalars of 5D
  SYM and should not be confused with the scalar fields for the linear
  sigma models considered in the previous sections.}  It is known that
in this case finding instanton-soliton solutions requires turning on
an electric field, which stabilizes the \emph{classical} size
\cite{Lambert:1999ua,Peeters:2001np}. From the point of view of the
quantum theory, turning on an electric field generates a potential on
the centered moduli space,
\begin{equation}\label{potential}
\delta {\rm v}(U^m) = \frac{2 \pi^2}{g_{\rm YM}^2} \langle \Phi \rangle^2 \rho^2~,
\end{equation}
and lifts the flat direction associated with the instanton-soliton
size.  Although $\rho$ is no longer a true modulus, the VEV provides
an additional dimensionless parameter, $ \epsilon := g_{\rm YM}^2
\langle\Phi \rangle$, that can be adjusted so that we remain in the
small-potential approximation, where it is still
appropriate to represent states as $L^2$-wavefunctions on
$\tilde{\mathcal{M}}$. In order to determine the precise form of the
resulting $L^2$-wavefunctions, one would need to compute the
centered-moduli-space Hamiltonian $\tilde H_{\rm eff}$, appearing in
\eqref{fullL} and \eqref{laglag}

Our formalism has been general enough to accommodate such potentials
on moduli space. Thus, despite the classical stabilization, one must
still integrate over all of moduli space, which includes arbitrarily
small sizes. However, as we have already discussed, this means
treating the solitons semiclassically when $\rho\sim O (g_{\rm
  YM}^2)$, which is not valid because quantum corrections that have
been neglected become important. Hence, turning on the potential
\eqref{potential} does not enable one to salvage an argument for
faster-than-any-power suppression.

While none of these arguments definitively show that instanton-soliton
contributions are {\it not} suppressed compared to perturbative
processes, they at least allow for that possibility. Non-suppression
of the pair-production amplitude would provide a mechanism via which
the contribution of virtual soliton-antisoliton pairs to perturbative
processes can compete with the contribution from loops of perturbative
particles.  Such a mechanism is precisely what is called for in order
to avoid contradicting the assumption of finiteness: One would require
that the soliton-antisoliton contribution be divergent, with exactly
the right coefficient to cancel the divergence found in
\cite{Bern:2012di}.  This is an intriguing possibility, the
investigation of which would, however, require an alternative approach
to the one used here.


\bibliographystyle{utphys}   
\bibliography{proceedings}

\providecommand{\href}[2]{#2}\begingroup\raggedright\begin{thebibliography}{10}

\bibitem{Papageorgakis:2014dma}
C.~Papageorgakis and A.~B. Royston, ``{Revisiting Soliton Contributions to
  Perturbative Amplitudes},''
\href{http://arxiv.org/abs/1404.0016}{{\tt arXiv:1404.0016 [hep-th]}}.

\bibitem{Papageorgakis:2014jia}
C.~Papageorgakis and A.~B. Royston, ``{Scalar Soliton Quantization with Generic
  Moduli},'' \href{http://dx.doi.org/10.1007/JHEP06(2014)003}{{\em JHEP} {\bf
  1406} (2014)  003},
\href{http://arxiv.org/abs/1403.5017}{{\tt arXiv:1403.5017 [hep-th]}}.

\bibitem{Douglas:2010iu}
M.~R. Douglas, ``{On D=5 super Yang-Mills theory and (2,0) theory},''
  \href{http://dx.doi.org/10.1007/JHEP02(2011)011}{{\em JHEP} {\bf 1102} (2011)
   011}, \href{http://arxiv.org/abs/1012.2880}{{\tt arXiv:1012.2880 [hep-th]}}.

\bibitem{Lambert:2010iw}
N.~Lambert, C.~Papageorgakis, and M.~Schmidt-Sommerfeld, ``{M5-Branes,
  D4-Branes and Quantum 5D super-Yang-Mills},''
  \href{http://dx.doi.org/10.1007/JHEP01(2011)083}{{\em JHEP} {\bf 1101} (2011)
   083}, \href{http://arxiv.org/abs/1012.2882}{{\tt arXiv:1012.2882 [hep-th]}}.

\bibitem{Rozali:1997cb}
M.~Rozali, ``{Matrix theory and U duality in seven-dimensions},''
  \href{http://dx.doi.org/10.1016/S0370-2693(97)00361-4}{{\em Phys.Lett.} {\bf
  B400} (1997)  260--264},
\href{http://arxiv.org/abs/hep-th/9702136}{{\tt arXiv:hep-th/9702136
  [hep-th]}}.

\bibitem{Berkooz:1997cq}
M.~Berkooz, M.~Rozali, and N.~Seiberg, ``{Matrix description of M theory on
  T**4 and T**5},'' \href{http://dx.doi.org/10.1016/S0370-2693(97)00800-9}{{\em
  Phys.Lett.} {\bf B408} (1997)  105--110},
\href{http://arxiv.org/abs/hep-th/9704089}{{\tt arXiv:hep-th/9704089
  [hep-th]}}.

\bibitem{Witten:1995zh}
E.~Witten, ``{Some comments on string dynamics},''
\href{http://arxiv.org/abs/hep-th/9507121}{{\tt arXiv:hep-th/9507121
  [hep-th]}}.

\bibitem{Strominger:1995ac}
A.~Strominger, ``{Open p-branes},''
  \href{http://dx.doi.org/10.1016/0370-2693(96)00712-5}{{\em Phys.Lett.} {\bf
  B383} (1996)  44--47},
\href{http://arxiv.org/abs/hep-th/9512059}{{\tt arXiv:hep-th/9512059
  [hep-th]}}.

\bibitem{Howe:2002ui}
P.~Howe and K.~Stelle, ``{Supersymmetry counterterms revisited},''
  \href{http://dx.doi.org/10.1016/S0370-2693(02)03271-9}{{\em Phys.Lett.} {\bf
  B554} (2003)  190--196},
\href{http://arxiv.org/abs/hep-th/0211279}{{\tt arXiv:hep-th/0211279
  [hep-th]}}.

\bibitem{Bern:2012di}
Z.~Bern, J.~J. Carrasco, L.~J. Dixon, M.~R. Douglas, M.~von Hippel, {\em et
  al.}, ``{D = 5 maximally supersymmetric Yang-Mills theory diverges at six
  loops},'' \href{http://dx.doi.org/10.1103/PhysRevD.87.025018}{{\em Phys.Rev.}
  {\bf D87} (2013)  025018},
\href{http://arxiv.org/abs/1210.7709}{{\tt arXiv:1210.7709 [hep-th]}}.

\bibitem{Banks:2012dp}
T.~Banks, ``{Arguments Against a Finite N=8 Supergravity},''
\href{http://arxiv.org/abs/1205.5768}{{\tt arXiv:1205.5768 [hep-th]}}.

\bibitem{Goldstone:1974gf}
J.~Goldstone and R.~Jackiw, ``{Quantization of Nonlinear Waves},''
\href{http://dx.doi.org/10.1103/PhysRevD.11.1486}{{\em Phys.Rev.} {\bf D11}
  (1975)  1486--1498}.

\bibitem{MR2068924}
N.~Manton and P.~Sutcliffe,
  \href{http://dx.doi.org/10.1017/CBO9780511617034}{{\em Topological
  solitons}}.
\newblock Cambridge Monographs on Mathematical Physics. Cambridge University
  Press, Cambridge, 2004.
\newblock \url{http://dx.doi.org/10.1017/CBO9780511617034}.

\bibitem{Derrick:1964ww}
G.~Derrick, ``{Comments on nonlinear wave equations as models for elementary
  particles},''
\href{http://dx.doi.org/10.1063/1.1704233}{{\em J.Math.Phys.} {\bf 5} (1964)
  1252--1254}.

\bibitem{Gervais:1975pa}
J.-L. Gervais, A.~Jevicki, and B.~Sakita, ``{Perturbation Expansion Around
  Extended Particle States in Quantum Field Theory. 1.},''
\href{http://dx.doi.org/10.1103/PhysRevD.12.1038}{{\em Phys.Rev.} {\bf D12}
  (1975)  1038}.

\bibitem{Dorey:1993by}
N.~Dorey, M.~P. Mattis, and J.~Hughes, ``{Soliton quantization and internal
  symmetry},'' \href{http://dx.doi.org/10.1103/PhysRevD.49.3598}{{\em
  Phys.Rev.} {\bf D49} (1994)  3598--3611},
\href{http://arxiv.org/abs/hep-th/9309018}{{\tt arXiv:hep-th/9309018
  [hep-th]}}.

\bibitem{Ji:1991ff}
X.-D. Ji, ``{A Relativistic skyrmion and its form-factors},''
\href{http://dx.doi.org/10.1016/0370-2693(91)91185-X}{{\em Phys.Lett.} {\bf
  B254} (1991)  456--461}.

\bibitem{Lambert:1999ua}
N.~D. Lambert and D.~Tong, ``{Dyonic instantons in five-dimensional gauge
  theories},'' \href{http://dx.doi.org/10.1016/S0370-2693(99)00894-1}{{\em
  Phys. Lett.} {\bf B462} (1999)  89--94},
\href{http://arxiv.org/abs/hep-th/9907014}{{\tt arXiv:hep-th/9907014}}.

\bibitem{Peeters:2001np}
K.~Peeters and M.~Zamaklar, ``{Motion on moduli spaces with potentials},''
  \href{http://dx.doi.org/10.1088/1126-6708/2001/12/032}{{\em JHEP} {\bf 0112}
  (2001)  032},
\href{http://arxiv.org/abs/hep-th/0107164}{{\tt arXiv:hep-th/0107164
  [hep-th]}}.

\end{thebibliography}\endgroup


\end{document}